\documentstyle{mn}

\newif\ifAMStwofonts

\ifoldfss
  \ifCUPmtlplainloaded \else
    \NewTextAlphabet{textbfit} {cmbxti10} {}
    \NewTextAlphabet{textbfss} {cmssbx10} {}
    \NewMathAlphabet{mathbfit} {cmbxti10} {} 
    \NewMathAlphabet{mathbfss} {cmssbx10} {} 
  \fi
  \ifAMStwofonts
    \ifCUPmtlplainloaded \else
      \NewSymbolFont{upmath} {eurm10}
      \NewSymbolFont{AMSa} {msam10}
      \NewMathSymbol{\upi}     {0}{upmath}{19}
      \NewMathSymbol{\umu}     {0}{upmath}{16}
      \NewMathSymbol{\upartial}{0}{upmath}{40}
      \NewMathSymbol{\leqslant}{3}{AMSa}{36}
      \NewMathSymbol{\geqslant}{3}{AMSa}{3E}

      \let\leq=\leqslant \let\le=\leqslant
       \let\ge=\geqslant
    \fi
  \fi
\fi 

\ifnfssone
  \newmathalphabet{\mathit}
  \addtoversion{normal}{\mathit}{cmr}{m}{it}
  \addtoversion{bold}{\mathit}{cmr}{bx}{it}
  \newmathalphabet{\mathbfit} 
  \addtoversion{normal}{\mathbfit}{cmr}{bx}{it}
  \addtoversion{bold}{\mathbfit}{cmr}{bx}{it}
  \newmathalphabet{\mathbfss} 
  \addtoversion{normal}{\mathbfss}{cmss}{bx}{n}
  \addtoversion{bold}{\mathbfss}{cmss}{bx}{n}
  \ifAMStwofonts
    \ifCUPmtlplainloaded \else
      \UseAMStwoboldmath
      \makeatletter
      \new@mathgroup\upmath@group
      \define@mathgroup\mv@normal\upmath@group{eur}{m}{n}
      \define@mathgroup\mv@bold\upmath@group{eur}{b}{n}
      \edef\UPM{\hexnumber\upmath@group}
      \new@mathgroup\amsa@group
      \define@mathgroup\mv@normal\amsa@group{msa}{m}{n}
      \define@mathgroup\mv@bold\amsa@group{msa}{m}{n}
      \edef\AMSa{\hexnumber\amsa@group}
      \makeatother
      \mathchardef\upi="0\UPM19
      \mathchardef\umu="0\UPM16
      \mathchardef\upartial="0\UPM40
      \mathchardef\leqslant="3\AMSa36
      \mathchardef\geqslant="3\AMSa3E

      \let\leq=\leqslant \let\le=\leqslant
       \let\ge=\geqslant
    \fi
  \fi
\fi 

\ifnfsstwo
  \DeclareMathAlphabet{\mathbfit}{OT1}{cmr}{bx}{it}
  \SetMathAlphabet\mathbfit{bold}{OT1}{cmr}{bx}{it}
  \DeclareMathAlphabet{\mathbfss}{OT1}{cmss}{bx}{n}
  \SetMathAlphabet\mathbfss{bold}{OT1}{cmss}{bx}{n}
  \ifAMStwofonts
    \ifCUPmtlplainloaded \else
      \DeclareSymbolFont{UPM}{U}{eur}{m}{n}
      \SetSymbolFont{UPM}{bold}{U}{eur}{b}{n}
      \DeclareSymbolFont{AMSa}{U}{msa}{m}{n}
      \DeclareMathSymbol{\upi}{0}{UPM}{"19}
      \DeclareMathSymbol{\umu}{0}{UPM}{"16}
      \DeclareMathSymbol{\upartial}{0}{UPM}{"40}
      \DeclareMathSymbol{\leqslant}{3}{AMSa}{"36}
      \DeclareMathSymbol{\geqslant}{3}{AMSa}{"3E}

      \let\leq=\leqslant \let\le=\leqslant
       \let\ge=\geqslant
    \fi
  \fi
\fi 

\ifCUPmtlplainloaded \else
  \ifAMStwofonts \else 
    \def\upi{\pi}
    \def\umu{\mu}
    \def\upartial{\partial}
  \fi
\fi

\title[Starcounts in the Hubble Deep Field \& Galactic Structure Models]
  {Starcounts in the Hubble Deep Field: Constraining Galactic
Structure Models\thanks{Based on observations obtained with the
NASA/ESA Hubble Space Telescope, obtained at the Space Telescope
Science Instittute, which is operated by AURA, Inc., under NASA
contract NAS5-26555.}}
\author[R. A. M\'endez et al.]
  {R. A. M\'endez,$^1$ D. Minniti,$^{2,1}$ G. De Marchi,$^1$ A.
Baker,$^{3,4}$ and W. J. Couch$^{3,5}$\\
  $^1$European Southern Observatory, Karl-Schwarzschild-Stra$\beta$e
2, D-85748, Garching b. M\"unchen, Germany\\
  $^2$Lawrence Livermore National Laboratory, MS L-413, P.O. Box 808,
Livermore CA 94550, USA\\
  $^3$European Southern Observatory, Visiting Scientist\\
  $^4$Institute of Astronomy, Madingley Road, Cambridge, CB30HA, England\\
  $^5$School of Physics, University of New South Wales, Sydney 2052, Australia}
\date{Accepted 1996 July 22. Received 1996 April 21}
\pagerange{\pageref{firstpage}--\pageref{lastpage}}
\pubyear{1996}

\def\LaTeX{L\kern-.36em\raise.3ex\hbox{a}\kern-.15em
    T\kern-.1667em\lower.7ex\hbox{E}\kern-.125emX}

\begin{document}
\label{firstpage}
\maketitle
\begin{abstract}
Stellar sources are identified in the Hubble Deep Field, and accurate
colours and magnitudes are presented. The predictions of a Galactic
starcounts model are compared with the faint stellar counts in this field.
The model reproduces the observations very well in the magnitude range 
$21.0 < V \le 26.4$, while it overpredicts the counts by a factor of four
in the range $26.4 < V \le 30.0$. The luminosity function for halo objects
must be a factor of two smaller than that predicted by an extrapolation
of the solar-neighborhood luminosity function for disc stars (with
95\% confidence level). This result, seen before in deep Hubble Space
Telescope images of globular clusters, is therefore confirmed for the halo field population. The possible nature of a group of faint-blue
objects is also investigated, concluding that they are most likely
non-stellar. The possibility that they are QSOs is ruled out. If we
insist upon their stellar nature, they would be halo white dwarfs, with either a
very steep halo white dwarf luminosity function for $M_v > +11.0$, or a stellar
density 0.4 times that of the disc white dwarfs in the solar-neighborhood.
\end{abstract}


\section{Introduction}

The use of starcount models to constrain {\it global} Galactic structure
parameters has proved to be an effective way of investigating the
broad properties of stellar populations in our Galaxy
(Reid and Majewski 1993).  However, studies of
the distribution at large distances above the plane require
accurate photometry extending to faint magnitudes ($V> 20$), and such
datasets are still rare (Majewski 1994, Robin 1994).

The faint end of the luminosity function (and
its concomitant implications for the amount and nature of the local
dark matter), is an important issue that can be resolved through 
starcounts to very faint magnitudes. A major obstacle in this area has
been the difficulty in distinguishing faint images of stars from
galaxies (which dominate the counts for $V> 20$), which has limited
most of the ground-based work to $V \leq 20$. Only observations
with a resolution higher than that imposed by the atmospheric seeing
will be able to address these issues.

Recently, the Hubble Space Telescope (HST) has devoted 150 orbits in 
the continuous viewing zone to observe a high-galactic 
latitude field. This observation reaches very faint magnitudes ($V
\approx 30$), and therefore can be used to
explore starcounts to very faint magnitudes at the best resolution
available. The observations in this so-called Hubble Deep Field 
(HDF, Williams et al. 1996) are analyzed in this article in the 
context of a galactic
structure model. Constraints on the level of the halo
luminosity function, and the nature of a population
of faint-blue point-like objects detected in the frames is explored.

Sections 2 and 3 describe the assembly of the catalogue of
stellar objects in the HDF. Section 4 compares the model predictions
with the observed counts. Section 5 summarizes our main conclusions.

\section{The HDF catalogue of sources}

The HDF object catalogue was created by running SExtractor
(Bertin 1995) on the combined F606W and F814W frames, totalling 70
hours of exposure time. This combined frame provides the deepest exposure for detections
at faint levels. Details of the catalogue creation can be found in
Clements and Couch (1996). From this catalogue, a list of point-like
objects was created by using
the neural-network classifier within
SExtractor that gives the probability of an object being point-like 
(in what follows this probability will be referred to as
the ``CLASS'' of an object, with $CLASS=0$ being an extended source
and $CLASS=1$ being a point-like object). This object
selection is quite different
from that presented by Flynn et al. (1996) and Elson et al. (1996) for
the HDF data, in that we have used a non-linear, neural-network, 
multi-parameter classifier fed with basic image parameters. Despite
the different
selection criteria, our sample is almost identical to that of Flynn et
al. (1996), judging from the similarity of the colour-magnitude
diagrams. The reliability of the classifier has been demonstrated by 
Bertin (1994), and Bertin \& Arnouts (1996).

Visual inspection of a subset of objects indicated that all objects with
$CLASS< 0.85$ were clearly extended. This was used to create a list of
sixteen possible stellar candidates for further inspection. SExtractor also
measured magnitudes from the F300W, F450W, F606W, and F814W frames
(however, most objects were too faint on F300W to be detected, see
Table 1).

\section{The catalogue of stars}

Figure 1 shows the colour-magnitude and colour-colour diagrams for our
sample of point-like objects. There
appear two clumps of points, one resembling a faint main-sequence
extension to low-luminosity stars, and a clump of blue, faint objects
(these objects have also been reported by Flynn et al. 1996, their
Figure 2). Figure 1a indicates the expected sequence of
Population I main-sequence stars of different spectral type as seen
through the HST filters. The blue objects mentioned before clearly depart from this sequence. Since
all of these faint-blue objects had $CLASS=0.85$ we suspected that
they could be barely resolved galaxies. 
In order to test this hypothesis we created
a histogram of the CLASS values to examine the distribution of
classes. We found (Figure 2) that there is
a broad local maximum at $CLASS=0.80$ with wings extending to
$CLASS=0.85$ which suggested that the blue objects belonged to the
same type of objects as the ones near this local maximum. We confirmed
this by plotting the colour-magnitude and colour-colour diagrams for
objects with $0.75 \le CLASS < 0.85$ (all of which are extended
objects 
as seen directly from the frames). These plots showed that
the extended objects fall in the same region as the faint-blue
objects found in our point-like sample. This suggests, therefore, that
these objects are non-stellar. As a
subgroup, these objects deserve special attention: We found that they 
{\it could not} be QSOs. Figures 1b-d indicate the locus of
QSOs in the colour-magnitude and colour-colour diagrams from recent
models by Baker (1995) for a range
of possible QSO spectral indices and emission-line strengths.
We also found (see next Section) that these objects could not be
white dwarfs (WDs), unless the halo WD luminosity function is 
extremely steep.

When creating our point-like sample we found that a plot of CLASS {\it
vs.} magnitude is not very informative because the overwhelming majority of
detections are galaxies, not stars. With only 14 point-like objects
and nearly 1500 galaxies spanning nearly 8 magnitudes of dynamic
range, a plot like this does not indicate the reliability of the
star-galaxy separation, nor it gives any information about the
completeness of our stellar sample. We have found that a much better
representation of the reliability of our star-galaxy separation is
that given by the histogram of separations, shown in Figure 2. The
validity of this procedure is confirmed by our numerical simulations
(Section 4.1) and by two other independent results, namely those of
Flynn et al. (1996) and Elson et al. (1996), as discussed below.

\begin{figure*}
\vspace{12.0cm}
\caption{Colour-colour and colour-magnitude diagrams in HST-STMAG
instrumental magnitudes for all objects with $CLASS\ge 0.85$. Panel a)
indicates the stellar sequence for spectral types A0 (bluest) to M8 (reddest).
The bluest object at the bottom of the stellar sequence is saturated and
not an A0 star. Panels b) to d) indicate the allowed range for the locus of
QSOs depending on their power-law spectral indices and the strength of
their emission lines. The locus has been computed from redshifts of
$z=0.1$ (bluest) to $z=5.0$ (reddest).}
\label{fig-1}
\end{figure*}

\begin{figure*}
\vspace{12.0cm}
\caption{Distribution of classes. The distribution of classes reveals a
clear separation between the local broad maximum at $CLASS= 0.80$
(extended objects) and a local peak at $CLASS= 0.94$ (point-like
objects) in the form of a valley with very few objects in the range $0.88 \le
CLASS < 0.92$, indicating a good star-galaxy separation.}
\label{fig-2}
\end{figure*}

Additionally, the two brightest objects in our point-like sample 
were also excluded
from the stellar list as these stars were saturated. Therefore, our
stellar sample consists of eight objects in a field of view of 
4.69 arc-min$^2$. Table 1 indicates the HST-STMAG instrumental magnitudes for
both the stellar sample and the faint-blue point-like objects.

\section{Starcounts modeling}

Colour-magnitude and colour-colour plots in the Johnson-Cousins system
are shown in Figure 3, where the faint-blue objects thought to be most
likely extragalactic have also been included. Magnitude limits (for a
$5 \sigma$ detection of galaxies) have been determined to be 30.3, and
29.0 for the V and I filters, respectively (these magnitude limits are
independent of colour in the range $1.8 \le V-I < 3.0$). We should
emphasize that we have {\it not} used colours at all to select our
stellar sample. Our calibrated photometry is also presented in Table 1.

\begin{table*}
\begin{minipage}{130mm}
\caption{HST-STMAG and Johnson-Cousins photometry for our point-like sample.}
\label{symbols}
\begin{tabular}{@{}cccccccccccccc}
\hline
\hline
ID & Chip & X & Y & F300W & F450W & F606W & F814W & V & B-V & V-I \\
\hline
1 & 2 & 1919.59 & 1913.17 & 23.610 &  21.717 & 21.528 & 21.715 & 21.11 &  1.05 &  0.70\\
2 & 3 & 1220.69 & 507.11  & 25.319 &  22.795 & 22.188 & 22.070 & 21.84 &  1.49 &  1.10\\
3 & 4 &  741.57 & 600.13  & ------ &  25.156 & 24.137 & 23.252 & 24.00 &  1.79 &  2.10\\
4 & 3 & 1055.63 & 1056.10 & ------ &  25.722 & 24.516 & 23.561 & 24.40 &  2.00 &  2.19\\
5 & 3 & 2003.79 & 1188.00 & ------ &  25.807 & 24.743 & 23.358 & 24.74 &  1.71 &  2.74\\
6 & 4 & 379.73  & 597.36  &  ------ &  25.916 & 25.228 & 25.146 & 24.87 &  1.60 &  1.05\\
7 & 4 & 409.84  & 1580.44 &  ------ &  26.858 & 25.652 & 24.951 & 25.46 &  2.07 &  1.86\\
8 & 2 & 1950.33 & 839.84  & 24.355 &  25.205 & 25.944 & 26.349 & 25.47 & -0.03 &  0.41\\
9 & 2 &  976.68 & 1272.12 & ------ &  27.447 & 26.530 & 26.234 & 26.23 &  1.82 &  1.33\\
10 & 2 & 906.05 & 1694.03 & ------ & 26.123 & 26.926 & 27.549 & 26.40 & -0.06 &  0.12\\
11 & 2 & 626.77 & 582.97  & ------ & 27.637 & 28.166 & 28.492 & 27.71 &  0.21 &  0.51\\
12 & 3 & 1026.22 & 813.84 & ------ & 27.498 & 28.106 & 27.973 & 27.77 & -0.01 &  1.12\\
13 & 4 & 1789.74 & 1306.06& ------ & 28.013 & 28.458 & 28.762 & 28.01 &  0.30 &  0.54\\
14 & 2 & 625.10  & 248.51  & ------ & 27.877 & 28.801 & 28.638 & 28.47 & -0.41 &  1.16\\
\hline
\hline
\end{tabular}
Chip refers to the WFPC2 chip in which the object appears, while X
and Y are the drizzled coordinates of the object (in pixels) on the
respective Chip.
\end{minipage}
\end{table*}

The paucity of objects fainter than $V \approx 26.2$ ($I \approx 24.4$)
(well above the magnitude
limit of our frames) is clearly seen in Figure 3a. This result is
consistent with the findings by Flynn et al. (1996). We depart from
their analysis in that we first {\it fit} a galactic structure model to the
bright data, showing consistency with the model, and {\it then} we
extrapolate to fainter magnitudes into the region where we do not 
observe any stars, to place constraints on the halo main-sequence 
luminosity function.

\begin{figure*}
\vspace{12.0cm}
\caption{Colour-magnitude (panels (a) to (c)) and colour-colour (panel
(d)) diagrams in the
Johnson-Cousins system. The STMAG instrumental magnitudes were converted
to the Johnson-Cousins system using the relationships by Holtzman et
al. (1995). Since these conversions depend on colour, an iterative
procedure was used to solve, simultaneously, for magnitudes in the
three passbands B, V, and I.} \label{fig-3}
\end{figure*}

We use a galactic structure model that incorporates the three major
contributors to the stellar counts in the solar neighborhood; a disc,
a thick-disc, and a halo. Details of the model can be found in
M\'endez and van Altena (1996). For the most important parameter
in this discussion, namely, the adopted halo luminosity
function, we have used a M3-like luminosity function. This function has
been constructed by padding the function of Sandage (1957) at the 
bright end ($M_v \le 3.4$) with the Paez et
al. (1990) function for fainter magnitudes. Beyond the Paez et al. 
completeness
limit ($M_v \ge 7.4$) we have adopted the disc luminosity function for
the solar neighborhood from Wielen et al. (1983). The function was
scaled to a density of 0.15\% of the stellar density at the solar 
neighborhood. The composite luminosity function has a ratio
of $2.5:23:65$ at $M_v=3.0:4.5:6.5$, similar to the best halo
representation obtained by Reid and Majewski (1993).

We have run the model in the magnitude range $21.0 \le V < 26.4$ and
$B-V> 1.0$ to match the colour and magnitude range of the observed
counts (Figure 3a). The number of stars predicted by the model in the
HDF field of view is 8.1 stars, while the observed number is 8. It should
be emphasized that {\it no scaling whatsoever} has been applied to
the model predictions, which are based purely on local values for
the stellar density and local normalizations. We have found that
variations in {\it all} of the model parameters (with the exception of
the luminosity function) within their observational uncertainties have
an effect smaller than the Poisson error in the observed counts (these
parameters, e.g., scale-heights,
scale-lengths, axial-ratio for the halo, etc., enter in a non-linear
way into the model predictions).This implies
that the predicted counts {\it are not} sensitive to the exact value
adopted for these parameters and that any significant discrepancy
between the model predictions and the observed value {\it has} to be
attributed to the only remaining free parameter, namely, the
luminosity function.

\subsection{The halo main-sequence luminosity function}

Figure 3a shows that in the range $26.4 \le V < 30.0$\, and $B-V
\ge +1.5$) there are no
observed objects. At $V=30.0$ we are still 0.3 
magnitudes above the $5 \sigma$ magnitude limit (which pertains to
galaxies with 16
connected pixels; the magnitude limit for point-like objects will be
correspondingly fainter), and we expect to
be fairly complete at this magnitude. Simulated images added to the HDF field
indicate that the star-galaxy separation software is reliable to
$V\approx I \approx 27.5$ and that the completeness at $V=30$ is
approximately 97\% (L. Yan, private communication). The results of the
extensive image simulations carried-out in the course of this
investigation will be presented elsewhere (Reid et al. 1996). Here we
only note that the question of misclassification is in our favor: since
the overwhelming majority of objects in the HDF field are galaxies, any misclassification will
likely bring more galaxies into this magnitude and colour range than
the number of stars misclassified as galaxies. Diffuse and faint
features associated with extended objects will not appear at the
faintest magnitude levels, thus forcing the software to classify them
as stars. This point has also been stressed by Elson et al. (1996), Section
2, and is clearly exemplified in their Figure 3. Since distinguishing 
point-like from extended sources requires approximately five times more
photons than just detection (Flynn et al. 1996, Section 2), the above
limit for reliable classification does imply that we can go
much fainter than that for detection (as it is indeed found to be the 
case from our simulations). The basic point here is that, even if the
classifier fails at $V\approx I \approx 27.5$, the non-detection of
point-like objects fainter than that implies a true lack of stellar
objects in this magnitude range (as long as we are above the
completeness limit). In this sense, the negative detection of
point-like sources in this magnitude range is an {\it absolute upper
limit} to the number of stars observed. Our high completeness at $V=
30$ (which is approximately of 97 \%) is not 
surprising in view that the magnitude limit ($5 \sigma$) for extended
sources is 30.3 (as pointed out previously), implying
that the corresponding $3 \sigma$ limit for point-like detections is
approximately $V= 31$. The magnitude limits for galaxies that we have
quoted above have been provided by the STScI-HDF team.

The model predicts 6
stars in the range $26.4 \le V < 30.0$\, and $B-V \ge +1.5$, including
0.3 disc stars, 1.6 thick-disc stars, and 4.1 halo stars. The number
of disc and thick-disc stars is formally consistent with no stars at
all. The number of predicted halo stars is,
however, larger than observed. This
suggests that the true halo field luminosity function is a factor
of four lower than implemented in the model. The mean distance
(computed self-consistently from the model) for these halo stars is in
the range 8.5 kpc (at $V=26.5$) to 15.2 kpc (at $V=29.9$). Therefore, 
the absolute magnitude range we
sample is $+11.9 \le M_v < +14.0$. In this magnitude range the halo
luminosity function has been padded with the Wielen et al. (1983)
function for the
solar neighborhood, which reaches a maximum at, 
precisely, $M_v= +13.0$. We thus conclude that, at the 95\% confidence
interval, the halo field luminosity
function is shallower, by a factor of two, than the Wielen function in this
magnitude range, and that, most likely, it is a factor of four
smaller. The small sample size, however, implies that this last
statement is only a $2 \sigma$ result, and that bigger samples would be
needed to place stronger constraints. Also, with the present data we
can not put constraints on the halo luminosity function at magnitudes 
fainter than $M_v= +14$. Our conclusion regarding the halo luminosity
function coincides with recent findings by HST on
the luminosity function of globular clusters (Paresce et al. 1995, De
Marchi \& Paresce 1995a,b). We should emphasize that, since the model
already predicts 3.4 halo stars in the range $26.4 \le V < 29.4$\, and $B-V
\ge +1.5$, the conclusions presented here regarding the halo luminosity
function {\it are not} very sensitive to the exact magnitude limit of
our sample of point-like objects. This also implies that the exact
completeness fraction near the magnitude limit is less of a concern.

\subsection{The white dwarf luminosity function}

With the aid of the model we have explored whether the six faint-blue
objects seen in the colour-magnitude diagram (with $V \ge 25$,
$-0.5< B-V < 0.5$ and $0 < V-I < 1.2$)
could indeed be WDs. In the magnitude range $25.4\le V <27.8$ and $B-V
< 0.1$ we predict 15 objects per square degree, which would imply no stellar
objects in our field-of-view. If we still assume that the objects we
detect {\it are} WDs, it is instructive to see what this would
imply in terms of their luminosity function. In one square-degree we
would expect 1 disc WD, 2 thick-disc WDs, and 12 halo
WDs. Therefore, if these objects are WDs, they would
be halo WDs at distances of about 10 kpc from the sun.
Excluding the two faintest objects in Figure 3 which could be 
misclassified galaxies (as found from our simulated images, see
Section 4.1), we 
are left with 4 objects (note that at $V =27.8$ we are well above the magnitude
limit of our sample). This would imply that the halo WD 
luminosity function would have to be a factor of 260 times that of the
disc (after normalizing to the density of halo stars in the solar
neighborhood), or 0.4 times the local density of (disc) WDs.
The mass density locked in halo WDs 
would be about $1.2 \times 10^{-4} M_{\sun}/pc^3$,
which is, nevertheless, smaller by two orders of magnitude than the density
required to explain the local circular velocity in the disc. Alcock et
al. (1996) have recently suggested that a fraction of the dark halo
could be indeed composed of halo WDs; the inferred masses from
the latest microlensing events observed in the MACHO program towards
the Large Magellanic Clouds 
are consistent with those of WDs. The alternative to just scaling
the disc WD luminosity function would be to assume a very steep WD
halo function for $M_v \ge +11.0$.

Besides the suggestion that this objects could indeed be unresolved
galaxies (Section 3), another apparent difficulty with the WD 
scenario are the colours of these objects which differ by
about 0.2 magnitudes from theoretical colours
of Hydrogen and Helium WDs produced recently by Bergeron et al.
(1995).

\section{Conclusions}

We find that a simple galactic structure model is able to reproduce
the main-sequence starcounts to $V \approx 26.5$. There is
an apparent lack of stars at fainter magnitudes, implying a halo
field luminosity function flatter by, at least, a factor of two than the solar
neighborhood Population I luminosity function in the range $+12 \le
M_v < +14.0$.

We also call attention to a group of faint and blue point-like
sources. Their nature is most easily explained as being
unresolved extragalactic objects (but not QSOs). On the other hand, if
they are assumed to be stellar
objects, they would most likely be halo WDs, implying a very
steep luminosity function for these objects.

The papers by Elson et al. (1996), Flynn et al. (1996), and ours are
three independent approaches to the study of the stellar sources in
the Hubble Deep Field. There are similarities, dictated by the use of the same data
set, but at the same time there are some important differences. We
believe our paper is more quantitative in nature, using a recent
Galactic starcount model (M\'endez and van Altena 1996), and therefore
it emphasizes the potential of these studies in the field of Galactic
Structure and Stellar Populations.

In the paper by Flynn et al. (1996), they do not show that their
bright stellar counts are in agreement with any Galactic structure
model, which we do. This is an important step because it both
validates the model, and permits a meaningful extrapolation to fainter
magnitudes, which is important for establishing constraints on the
halo luminosity function, as done in Section 4. On the
other hand, their paper does not provide any light on the nature of
the faint-blue objects which we also found. Their only remark about 
these extremely compact objects is that they would make it difficult to
search for faint blue stars, and that they would investigate their
possible nature. Instead, our analysis has shown that they could
certainly not be QSOs, and that they are most likely not stars either.
We have fully explored, though, the quantitative consequences of
assuming that they are stars. We provide estimates of the effect of
them being halo stars on the slope of the halo WD luminosity function,
and the overall stellar and mass density of this objects in the solar
vicinity. We have pointed out the resemblance of their CMD to ours,
but we can not proceed any further in this comparison, since their
paper does not contain a table indicating the photometric values for
their stellar sample.

	The analysis of the HDF starcounts in Elson et al. (1996) is
more along the lines of our own analysis, with the following important
differences:
\begin{enumerate}
\item The Halo luminosity function: They compare the cumulative observed
starcounts with models down to $V= 30$. We use the same magnitude cutoff
based on our completeness tests, but we additionally break the sample
into a `bright' portion (where we do observe stars) and a faint
portion (where we do not observe any stars). This separation of the
sample into two subsamples is important as it permits different
regions of the halo luminosity function to be mapped. For the bright
sample the match to the model is perfect while for the faint sample
there are discrepancies which lead to important constraints for the
halo luminosity function as described in Section 4.1. We note also that
Elson et al. have created their catalogue by using the F606W frames
only, while we have created our list from the co-added F606W and F814W
frames, allowing us to go deeper in terms of faint-level detections.

\item The faint-blue objects: Elson et al. discard the possibility that
they are halo WDs mainly because since there are no such objects
brighter than  V= 26, these putative halo WDs would have to be located
in a narrow shell at a distance of approximately 10 kpc from the Sun;
certainly an unphysical solution. From our model, the distance
distribution of these objects (should they be halo WDs) spans
the distance range 7.8 kpc at V= 25.5 to 19.3 kpc at V= 27.7. They are
not distributed in a thin shell, so this does not appear to be a good
reason for discarding this model. In addition, as pointed out in
Section 4.1, the fact that we do not observe blue objects brighter than $V
\approx 26$ could actually imply a steep halo WD luminosity function which is
not an unreasonable proposition in account of the older (and therefore
cooler and redder) nature of the halo WDs. We provide also specific
estimates to the corresponding mass density of halo WDs if these
faint-blue objects are regarded as such.

Elson et al. discuss QSOs in the general context of their
point-like sample by addressing mainly the expected {\it vs.} observed
density of QSOs as a function of magnitude. Our approach is
essentially complementary to theirs in that we actually compute the
expected locus of QSOs, which lead us to discard the possibility that
our faint-blue point like sample are QSOs. This, we believe, is an
important contribution to understanding the true nature of this objects.
\end{enumerate}

Summarizing, our analysis follows a logical sequence that
makes the most use of the observations to constrain models, departing
somewhat from the analysis by Flynn et al. and Elson et al., but
providing complementary information about the nature of the sample of
point-like objects found in the HDF. Despite our use of a completely
different method for selecting stellar candidates we obtain a similar
sample as that found by Elson et al. and Flynn et al.. This is quite
reassuring, specially considering the difficulties inherent to create
these samples at faint magnitude levels.

\section{Acknowledgements}

We have greatly benefited from conversations with Hans-Martin Adorf, 
Pierre Bergeron, Dave Clements, Stefan Dieters, Robert Fosbury, Markus
Kissler-Patig, Francesco Paresce, Alvio Renzini. We specially thank
Lin Yan for her contribution to the image simulations and the
corresponding limits on completeness and star-galaxy separation. We would also
like to thank the ST-ECF for arranging rapid access to the HDF data
for European Astronomers.

\section{Note added in proof}

Subsequent to acceptance of our paper, Flynn et al. (1996) have been
kind enough to provide us with their V and V-I photometric values. With
these we have performed a comparison of theirs and Elson et al. (1996)
photometry to ours. The results, for the objects in common
among the three studies, are shown in Table 2 below. We find (see Table 2) that the mean differences (in the sense others - this work)
are $\Delta V=0.17 \pm 0.10$ and $\Delta (V-I)=0.21 \pm 0.08$ with
respect to the photometry by Flynn et al. (1996), while the mean
differences for Elson et al. (1996) are $\Delta V=-0.13 \pm 0.14$,
$\Delta (V-I)=-0.24 \pm 0.15$, and $\Delta (B-V)=0.03 \pm 0.20$ (or
$\Delta (B-V)=-0.06 \pm 0.12$ excluding objects 10 and 12 which belong
to the possibly extended faint-blue sources). Our
photometry seems to be midway between that of these other works. The origin of
the discrepancy is unknown, but it can probably be traced back to the
different methods employed to transform from the HST instrumental
magnitudes to the Johnson-Cousins system. For example, Flynn et al.
have used the transformation described in Bahcall et al. (1994, ApJ,
435, L51) which makes use of spectrophotometric standards to determine
the zero-points and slopes of the conversion from the HST filters to
the Johnson-Cousins V and I filters for each of the wide-field
cameras. On the other hand, Elson et al. (1996) and ourselves use the
calibration provided by Holtzmann et. al (1995). This explains why the
difference in V and B-V between the photometry by Elson et. al and our work is
negligible within the error bars. However, the large discrepancy
in V-I is rather uncomfortable since the same transformations have been
employed.

\begin{table}
\caption{Photometric differences between Flynn et al. (1996), Elson
et. al (1996), and this work.}
\label{symbols}
\begin{tabular}{@{}cccccc}
\hline
\hline
ID & $\Delta V_{F}$ & $\Delta (V-I)_{F}$ & $\Delta V_{E}$ & $\Delta
(V-I)_{E}$ & $\Delta (B-V)_{E}$ \\
\hline
1 & 0.09   & 0.18 & ------& ------& ------ \\ 
2 & 0.27   & 0.19 & -0.01 & -0.13 & -0.12  \\
3 & 0.17   & 0.25 & -0.29 & -0.40 &  0.04  \\
4 & 0.35   & 0.26 & -0.15 & -0.29 & -0.07  \\
5 & 0.18   & 0.14 & -0.35 & -0.51 &  0.13  \\
6 & 0.16   & 0.27 & -0.02 & -0.13 & -0.16  \\
7 & 0.14   & 0.25 & -0.17 & -0.27 & -0.02  \\
9 &   0.19 & 0.31  & -0.17 & -0.16 & -0.20 \\
10 & -0.02 & 0.04  &  0.09 & -0.04 &  0.31 \\
12 & ------& ------& -0.11  & -0.22 &  0.35 \\
\hline
\hline
\end{tabular}
ID refers to our sequential number on Table 1. All the differences
($\Delta$) are in the sense others - this paper. The subscript `F'
refers to the differences Flynn et al. (1996) - this paper, while the
subscript `E' refers to the differences Elson et al. (1996) - this paper.
\end{table}

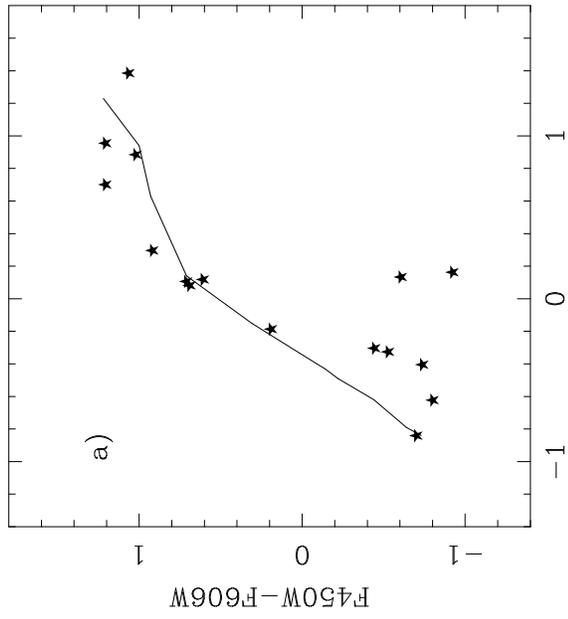
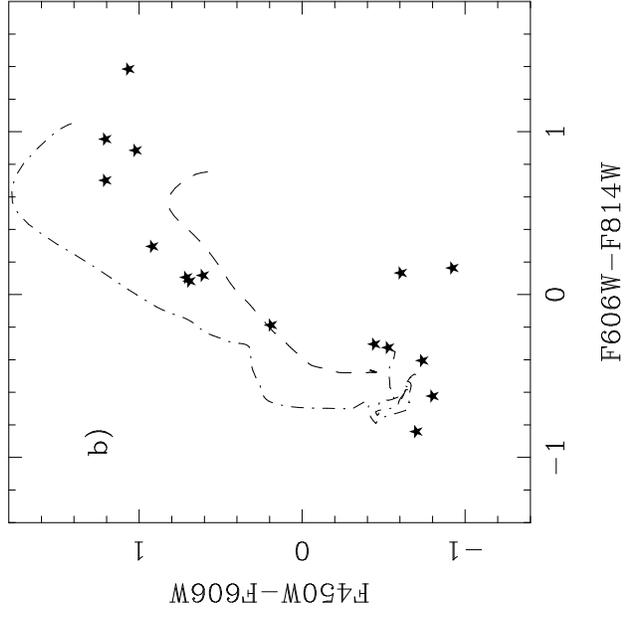
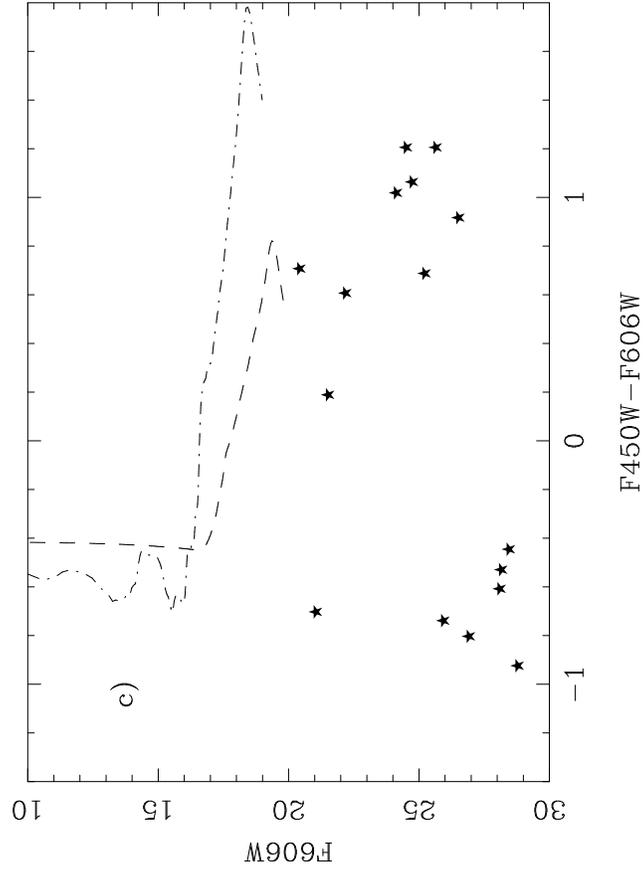
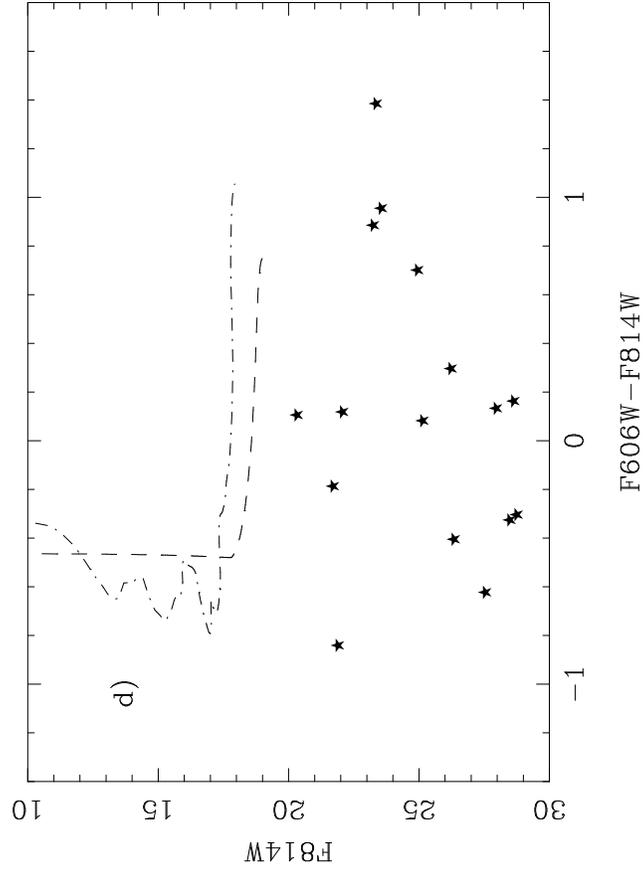
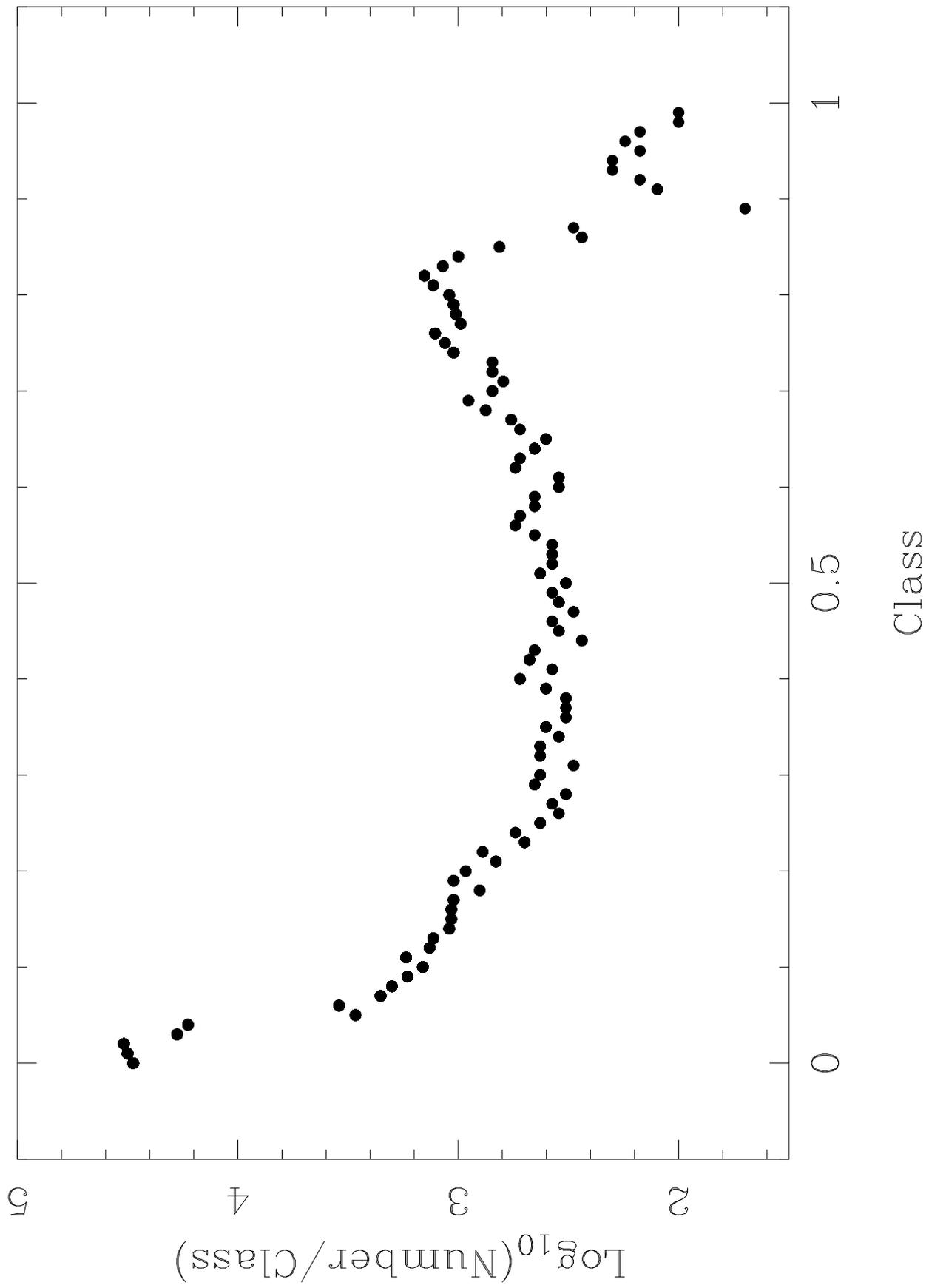
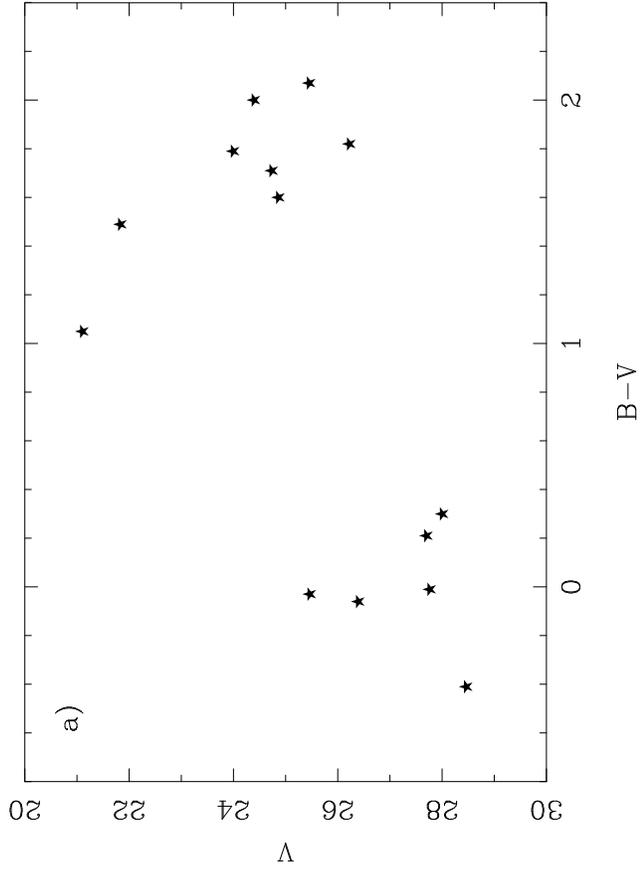
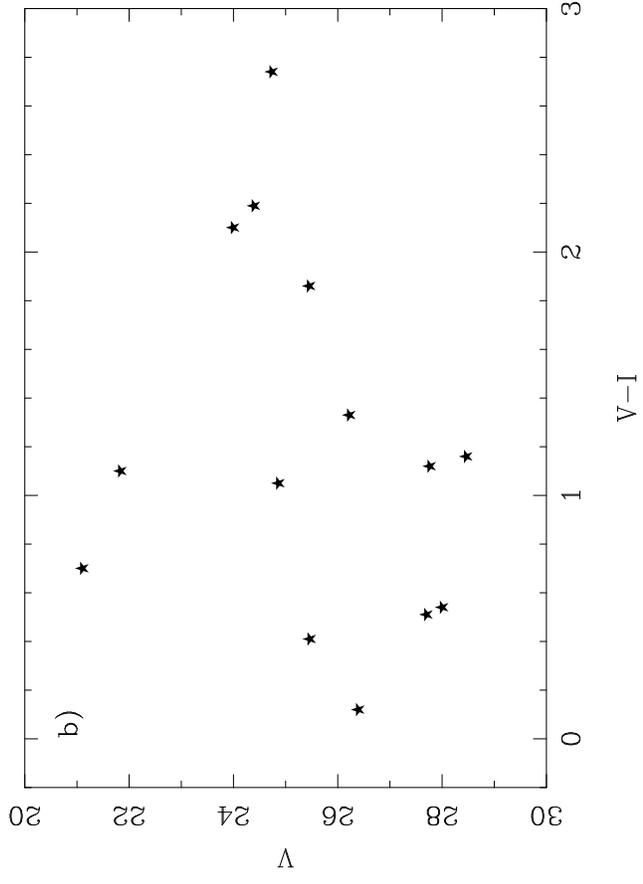
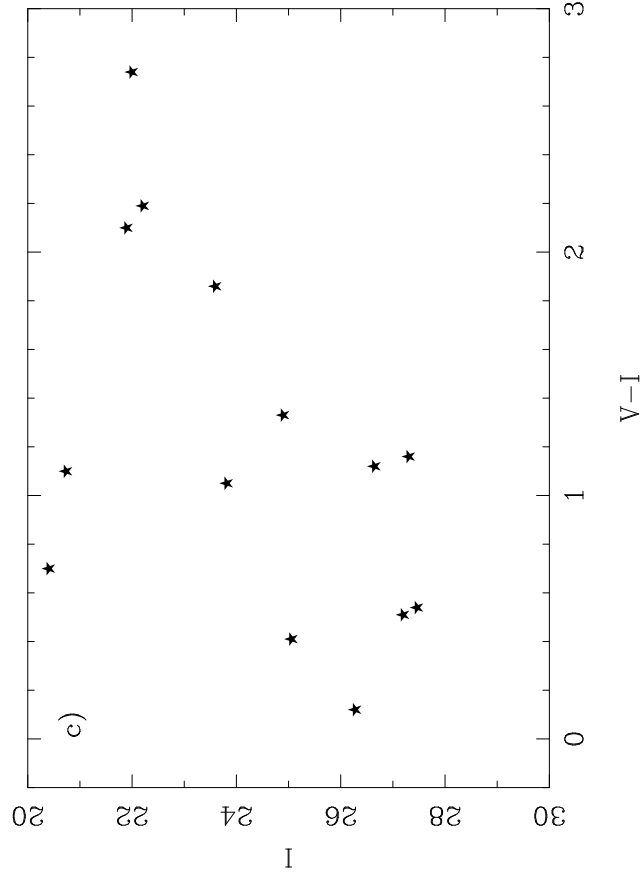
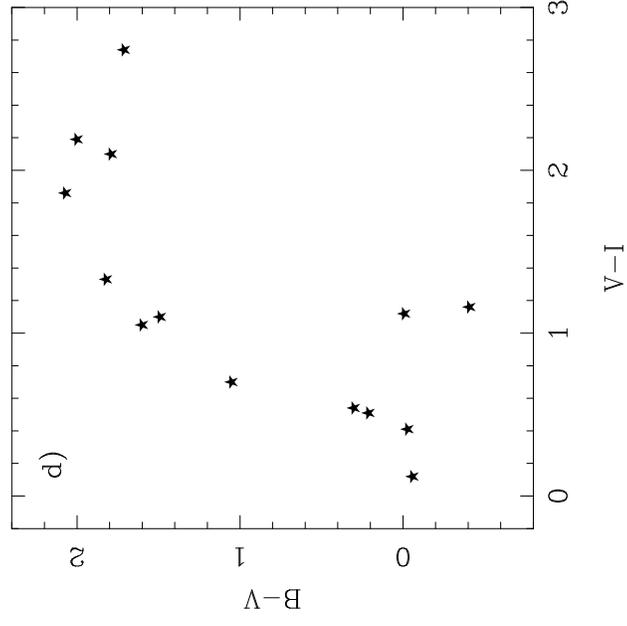
\end{document}